# How the alteration of a thin wall for S-S' di quark pairs signifies an Einstein constant dominated cosmology and a role for Extra dimensions in initial nucleation of a new universe

A. W. Beckwith

## ABSTRACT

We use as a model of how nucleation of a new universe occurs, assuming a di quark identification for S-S' constituent parts of a scalar field. We construct a model showing evolution from a dark matter dark energy mix to a pure cosmological constant cosmology due to changes in the slope of a graph of the resulting scalar field. The initial potential system employed is semi classical in nature, becoming non-classical at the end of chaotic inflation at the same time cosmological expansion is dominated by the Einstein cosmological constant. We use Scherrer's derivation of a sound speed being zero during initial inflationary cosmology — and change it afterwards as the slope of the scalar field moves away from a thin wall approximation. Furthermore, the results in Bo Qin's article about extra dimensions from dark matter, permit us to show the impact of dimensionality upon the role of semi classical approximations to inflation models .We conclude that the new force law specified by Bo Qin and additional dimensions would play a role in the early universe and be extremely important to the onset of inflationary expansion due to nucleation of di quark pairs in a S-S' configuration.

Correspondence: A. W. Beckwith:    projectbeckwith2@yahoo.com

## I. INTRODUCTION

Recently, Quin, Pen, and Silk[1] presented evidence as to how three extra dimensions play a role in explaining how at very short distances gravity would have a $r^{-5}$ spatial behavior dependence in force calculations. The emphasis given was with regards to general dark matter masses, giving them a specific upper bound. The new variant of force law was relevant for scales at or below $1nm$ in length. We find that this new force law and the additional dimensions would play a role in early universe nucleation models. This leads to very semi classical behavior at the onset of nucleation, and perhaps pertinent to the loss, initially, of a strict thin wall approximation to the domain walls for initial states of matter at the onset of inflationary cosmology. The initial additional dimensions, $n$, were specified as leading to a force for small distance scales below a crucial radius of $R$ leading to force with a spatial variance of $r^{-2-n}$, which we believe plays a crucial role in early universe nucleation models. Arguments they presented[1,2] so happened to fix this the value of $n$ as 3, which is enough to specify for very small dimensional settings a highly repulsive initial starting point for cosmological inflation, especially if the value of $R >> l_P$, with the initial radius of a nucleating universe being of the order of magnitude of $l_P$ at or before a Planck time $t_P$.

Our model shows that a semi-classical phase-state formed from initial di quark pairs in a region of the order of magnitude of Planck's constant for length $l_P$ changes to a physical system whose evolution is dominated by the Einstein cosmological constant. The initial phase state, which we approximate by a thin wall approximation, is similar to the semi-classical bounce state that Sidney Coleman[3] postulated; however, it changes in time to a very different system at the end of cosmological inflation. The model advantages are first that we provide a template for employing baryonic states to form dark matter as a driving force for the formation and expansion of cosmological states to the present conditions of our present universe. We also give initial conditions for the formation of CMB, which are not readily explained by current models. In addition, this model ties in with being able to use the Veneziano model of strength of all forces,[4] gravitational and gauge alike. Veneziano's model is one of the simplest ways to use Planck's length $l_P$ for an initial starting point for cosmological nucleation and expansion from the formation of di quark pairs with a very high number of degrees of freedom in a confined state.

## II. BRIEF RE CAP OF QINS EXTRA DIMENSIONS FROM DARK MATTER ARTICLE

As mentioned, Quinn's article[1] gives a new force law, with respect to distances at or below $1nm$ in length. As presented in the article, this appears to be a verification of the existence of small but non infinitesimal extra dimensions. The key assumption which was used in their paper was a force law of the general form for distances $r \ll R$:

$$F = \alpha \cdot \frac{GMm}{r^{2+n}} \qquad (1)$$

Here, $\alpha$ is a constant with dimensions $[length]^n$, G is the gravitational constant, and $M$ and $m$ are the masses of the two particles and. $\alpha \equiv R^n$ was set, while the value of $n$ was, partly to fit with an argument given by Volt and Wannier[2] that the quantum mechanical cross section for collision is twice the corresponding classical value, if one assumes a central force field dependence of $r^{-5}$ This all together, if one assumes that initially $r$ is of the order of magnitude of Planck's length $l_P$ would lead to extremely strong pressure values upon the domain walls of a nucleated scalar field initial states, which I claim would lead to a quite necessary collapse of the thin wall approximation. This collapse of the thin wall approximation set the stage for an Einstein constant dominated regime in inflation, if one adheres to a version of Scherrer's K essence theory[5] results for modeling the di quark pairs used as an initial starting point for soliton-anti soliton pairs(S-S') in the beginning of quantum nucleation of our universe.

### III. HOW TO ANALYZE PHYSICAL STATES IN THE PRECURSORS TO INFLATIONARY COSMOLOGY

Let us first consider an elementary definition of what constitutes a semi classical state. As visualized by Buniy and Hsu,[6] it is of the form $|a\rangle$ which has the following properties:

i) Assume $\langle a|1|a\rangle = 1$

(Where 1 is an assumed identity operator, such that $1|a\rangle = |a\rangle$)

ii) We assume that $|a\rangle$ is a state whose probability distribution is peaked about a central value, in a particular basis, defined by an operator $Z$

a) Our assumption above will naturally lead, ***for some*** $n$ values

$$\langle a|Z^n|a\rangle \equiv (\langle a|Z|a\rangle)^n \qquad (2)$$

Furthermore, this will lead to, if an operator $Z$ obeys Eq. (2) that if there exists another operator, call it $Y$ which does not obey Eq. (2), that usually we have non commutativity

$$[Y, Z] \neq 0 \qquad (3)$$

Buniy and Hsu[6] speculate that we can, in certain cases, approximate a semi classical evolution equation of state for physical evolution of cosmological states with respect to classical physics operators. This well may be possible for post inflationary cosmology; however, in the initial phases of quantum nucleation of a universe, it does not apply.

To review our model of S-S' pair nucleosynthesis for di quark pair states in an early universe, first is the issue of how the potential evolved. Namely:

$$\begin{aligned} V_1 & \to V_2 & \to V_3 \\ \phi(increase) \leq 2\cdot\pi & \to \phi(decrease) \leq 2\cdot\pi & \to \phi \approx \varepsilon^+ \\ t \leq t_P & \to t \geq t_P + \delta\cdot t & \to t \gg t_P \end{aligned} \tag{4}$$

We described the potentials $V_1$, $V_2$, and $V_3$ in terms of S-S' di quark pairs nucleating and then contributing to a chaotic inflationary scalar potential system.

$$V_1(\phi) = \frac{M_P^2}{2}\cdot(1-\cos(\phi)) + \frac{m^2}{2}\cdot(\phi-\phi^*)^2 \tag{4a}$$

$$V_2(\phi) \approx \frac{(1/2)\cdot m^2\phi^2}{(1+A\cdot\phi^3)} \tag{4b}$$

$$V_3(\phi) \approx (1/2)\cdot m^2\phi^2 \tag{4c}$$

Note that Eq. (3a) is a measure of the onset of quantum fluctuations[7]

$$\phi^* \equiv \left(\frac{3}{16\cdot\pi}\right)^{\frac{1}{4}} \cdot \frac{M_P^{3/2}}{m^{\frac{1}{2}}}\cdot M_P \to \left(\frac{3}{16\cdot\pi}\right)^{\frac{1}{4}}\cdot \frac{1}{m^{\frac{1}{2}}} \tag{4d}$$

and should be seen in the context of the fluctuations having an upper bound specified by[7]

$$\tilde{\phi}_0 > \sqrt{\frac{60}{2\cdot\pi}} M_P \approx 3.1 M_P \tag{4e}$$

Also, the fluctuations Guth[7] had in mind were modeled via

$$\phi \equiv \tilde{\phi}_0 - \frac{m}{\sqrt{12 \cdot \pi \cdot G}} \cdot t \tag{4f}$$

This is for his chaotic inflation model using his potential; which we call the third potential in Eq. (4c)

However, I show elswhere[8] that for the false vacuum hypothesis to hold for Eq. (4a) that there is

$$V_1(\phi_F) - V_1(\phi_T) \cong .373 \propto L^{-1} \cong \alpha \tag{4g}$$

Let us now view a toy problem involving use of a S-S' pair which we may write as[5]

$$\phi \equiv \pi \cdot [\tanh b(x - x_a) + \tanh b(x_b - x)] \tag{5}$$

We can, in this give an approximate wave function as given by:

$$\psi \cong c_1 \cdot \exp(-\tilde{\alpha} \cdot \phi(x)) \tag{6}$$

Then we can look to see if we have[6]

$$\left( \int_{x_a}^{x_b} \psi \cdot V_i \cdot \psi \cdot 4\pi \cdot x^2 \cdot dx \right)^N \equiv \int_{x_a}^{x_b} \psi \cdot [V_i]^N \cdot \psi \cdot 4 \cdot \pi \cdot x^2 \cdot dx \Bigg|_{i=1,2,3} \tag{7}$$

Please see the conclusion for misgivings I have about this very simplified model in Eq. (7). Eq. (7) would likely be redone substantially in a future calculation with brane world type of topological defects. Assuming that this is a valid initial dimensional approximation, we did the following for the three potentials.

a. Assumed that the scalar wave functional term was decreasing in '*height*' and increasing in '*width*' as we moved from the first to the third potentials. $\phi$ also had a definite evolution of the domain wall from a '*near perfect*' thin wall approximation to one which had a considerable slope existing with respect to the wall.

b. We also observed that in doing this sort of model that there was a diminishing of applicability of Eq. (7) for large *N* values, regardless if or not the thin wall approximation was weakened as we went from the first to the third potential system. In doing to, we also noted that even in Eq. (7) for the first potential, where Eq. (7) was almost identically the same values on both sides of the inequality, that Eq. (7) had diminishing applicability as a result for decreasing *b* values in Eq. (5), which corresponded to when the thin wall approximation was least adhered to.

We also observed that for the third potential, that there was never an overlap in value between the left and right hand sides of Eq. (7), regardless of whether the thin wall approximation was adhered to. In other words, the third potential was least linkable to a semi classical approximation of physical behavior linkable to a physical system, while Eq. (7) worked best for a thin domain wall approximation to Eq. (5) in the driven sine Gordon approximation of a potential system. In all this, we assumed that the small perturbing term added to the $(1-\cos(\phi))$ part of Eq. (7) was a physical driving term to a very classical potential system $(1-\cos(\phi))$ which had a quantum origin consistent with the interpretation of a false vacuum nucleation of the sort initially formulated by Sidney Coleman.[1] Furthermore, as we observed an expanding 'width' in Eq. (5), the alpha term in Eq. (6) shrank in its value, corresponding to a change in the position of constituent S-S' components in the scalar field given in this model. The S-S' terms roughly corresponded to di quark pairs.

c. Chaotic inflation in cosmology is, in the sense a quartic potential portrayed by Guth,[7] a general term for models of the very early Universe which involve a short period of extremely rapid (exponential) expansion; blowing the size of what is now the observable Universe up from a region far smaller than a proton to about the size of a grapefruit (or even bigger) in a small fraction of a second. This process smoothes out space-time to make the Universe flat, but is not in the model presented linkable in the chaotic inflationary region given by the third potential to any semi classical arguments. The relative good fit of Eq. (7) for the first potential is in itself an argument that the thin wall approximation breaks down past the point of baryogenesis after the chaotic inflationary regime is initiated by the third potential as modeled by Guth.[7]

To summarize the numerical procedures in the set of simulations for Eq. (7), they are: For the first potential, Eq. (4a), $\tilde{\alpha} \to .373$ in Eq. (6), and $b \to 20$ in Eq (5); Eq (7) gives us:

$$\left( \int_{-x_b}^{x_b} \psi \cdot V_1 \cdot \psi \cdot 4\pi \cdot x^2 \cdot dx \right)^{N=10} \equiv 5.49 \times \text{E-5} \tag{7a1}$$

while

$$\left( \int_{-x_b}^{x_b} \psi \cdot (V_1)^{10} \cdot \psi \cdot 4\pi \cdot x^2 \cdot dx \right) \equiv 5.92 \times \text{E-5} \tag{7a2}$$

This assumes that the second term in the first potential, Eq (3a), is 1/100 of the first term. Were we to have a smaller $b$ term, the relative overlap of Eq. (6a1) and Eq. (7a2) would go down, and it goes up with increasing $b$ values.

If we pick $A = .5$ in the second potential — Eq. (4b), $\tilde{\alpha} \to .373/2$ in Eq. (**6**), and $b \to 10$ in Eq. (5) — a halving of the height of the phase $\phi$ and a doubling of the 'length' integrated over Eq. (7) gives us:

$$\left( \int_{-2x_b}^{2x_b} \psi \cdot V_2 \cdot \psi \cdot 4\pi \cdot x^2 \cdot dx \right)^{N=10} \equiv 2.286 \times E - 8 \tag{7b1}$$

and

$$\left( \int_{-2x_b}^{2x_b} \psi \cdot (V_2)^{10} \cdot \psi \cdot 4\pi \cdot x^2 \cdot dx \right) \cong 0 \tag{7b2}$$

As with the first potential, the relative divergence of the left and right hand sides of Eq. (7) go up if $b$ gets smaller and decrease if $b$ gets larger. Still, this has a far less rigorous fit between the left and sides of Eq. (7) fit together than what happens with the first potential situation.

And, then, finally we have the chaotic inflationary potential given by Guth,[4] which shows no overlap at all in either side of Eq. (7). For the thrid potential, Eq. (3c), $\tilde{\alpha} \to .373/4$ in Eq. (6), $b \to 5$ in Eq. (5), and a division by 4 of the height of the phase $\phi$ and multiplication by four of the 'length' integrated over results in

$$\left( \int_{-4x_b}^{4x_b} \psi \cdot V_3 \cdot \psi \cdot 4\pi \cdot x^2 \cdot dx \right)^{N=10} \equiv 2.707 \times E - 11 \tag{7c1}$$

and

$$\left( \int_{-4x_b}^{4x_b} \psi \cdot (V_3)^{10} \cdot \psi \cdot 4\pi \cdot x^2 \cdot dx \right) \equiv 3.258 \times E + 10 \tag{7c2}$$

These results hold, even if b is increased in value. Namely, the overlap vanishes completely.

**Appendix I** offers even more striking results. Namely that if one uses a higher 6 dimensional 'volume' element for initial nucleated space, that the agreement of Eq. (7) for a spatial six dimensional analysis as a starting point for the first potential will lead to an almost exact equality. Furthermore, if we use a normalization procedure as outlined in that appendix, and compare the ratios of both sides, that the relative slope of the scalar field will not be terribly important, in determining the relative contributions to both sides of Eq. (7) for 2$^{nd}$ and 3$^{rd}$ potentials. Still though we argue that for especially the 1$^{st}$ potential that the higher dimensionality enhances the likelihood of a semi classical analysis being a good starting point, even though it appears to have only weak links to the chaotic inflationary 3$^{rd}$ potential as given in this analysis.

The comparison of the evolution of these different cases for Eq. (7) argue that if we show that in between the physical states represented from the first to the third potentials there is a phase change which has measurable consequences for cosmological evolution. Furthermore, we can employ a different paradigm as to how topological defects (kinks and anti kinks) contribute to the onset of initial conditions at the beginning of inflationary cosmology. Currently, as seen by Mark Trodden[9] and Trodden et al,[10] topological defects are similar to D branes of string theory; while this S-S' (soliton-antisoliton) construction permits extensions to super-symmetric theories, it obscures direct links to inflationary cosmological potentials such as Guth's[7] harmonic potential.

The zeroth level assumption underlying this is that there could be a C-P violation in the initial phases of states of matter. This in turn leads to Baryon matter state separation into Baryon-anti Baryon pairs (di quark pairs) which in turn would lead to the S-S' pair formation alluded to in Guth.[7] If the di quark pairs form, we would have a situation where an overall topological charge Q would tend to then vanish for our physical system.

To make the linkage clearer, we can present the di quark S-S' pairs as an initial starting point for times $t \leq t_P$, where $t_P$ is Planck's discretized smallest unit of time as a coarse graining of time stepping in cosmological evolution. Initially, let us look at work by Zhitnitsky[11] about formation of a soliton object via a so called di quark condensate.

**a)** A C-P violation in initial states would lead to an initial Baryon condensate of matter separating into actual S-S' di quark pairs:

**b)** for times less than or equal to Planck time $t_P$ the potential system for analyzing the nucleation of a universe is a driven Sine Gordon system,[12] with the driving force in magnitude far less than the overall classical Sine Gordon potential.

**c)** for this potential system, topological charges for a S-S' di quark pair stem prior to Planck time $t_P$ cancel out, leaving a potential proportional to $\phi^2$ minus a contribution due to quantum fluctuations of a scalar field being equal in magnitude to a classical system, with the remaining scalar potential field contributing to cosmic inflation in the history of the early universe.

The next assumption is that a vacuum fluctuation of energy equivalent to $\Delta t \cdot \Delta E = \hbar$ will lead to the nucleation of a new universe, provided that we are setting our initial time $t_P \approx \Delta t$ as the smallest amount of time which can be ascertained in a quantum universe.

If a phase transition occurs right after our nucleation of an initial state, it is due to the time of nucleation actually being less than (or equal to) Planck's minimum time interval $t_P$, with the length specified by reconciling the fate of the false vacuum potential used in nucleation with a Bogomol'nyi inequality specifying the vanishing of topological charge[13]. We can use S-S' di quark pairs to represent an initial scalar field, which, after time $t_P \approx \Delta t$, will descend into the typical chaotic inflationary potential used for inflationary cosmology.

## IV. INCLUDING IN NECESSARY AND SUFFICIENT CONDITIONS FOR FORMING A CONDENSATE STATE AT OR BEFORE PLANCK TIME $t_P$

For a template for the initial expansion of a scalar field leading to false vacuum inflationary dynamics in the expansion of the universe, Zhitnitsky's[14] formulation for how to form a condensate of a stable soliton style configuration of cold dark matter is a useful starting point for how an axion field can initiate forming a so called QCD ball. Zhitnitsky[14] uses quarks in a non-hadronic state of matter that, in the beginning, can be in di quark pairs. A di quark pair would permit making equivalence arguments to what is done with cooper pairs and a probabilistic representation as to find the relative 'size' of the cooper pair. We assume an analogous operation can be done with respect to di quark pairs. In doing so, calculations[14] for quarks being are squeezed by a so called QCD phase transition due to the violent collapse of an axion domain wall. The axion domain wall would be the squeezer to obtain a so called S-S' configuration. This presupposes a formation of a highly stable soliton type configuration in the onset due to the growth in baryon mass

$$M_B \approx B^{8/9} \tag{8}$$

This is due to a large baryon (quark) charge $B$ which Zhitnitsky[14] finds is smaller than an equivalent mass of a collection of free separated nucleons with the same charge. This provides a criteria for absolute stability by writing a region of stability for the QCD balls dependent upon the inequality occurring for $B. > B_C$ (a critical charge value)

$$m_N > \frac{\partial M_B}{\partial B} \tag{9}$$

He[14] furthermore states that stability, albeit not absolute stability is still guaranteed for the formation of meta stable states occurring with

$$1 << B < B_C \tag{10}$$

If we make the assumptions that there is a balance between Fermi pressure $P_f$ and a pressure due to surface tension, with $\sigma$ being an axion wall tension value[11] so that

$$\left(P_\sigma \cong \frac{2\sigma}{R}\right) \equiv \left(P_f \cong -\frac{\Omega}{V}\right) \tag{11}$$

This pre supposes that $\Omega$ is some sort of thermodynamic potential of a non interacting Fermi gas, so that one can then get a mean radius for a QCD ball at the moment of formation of the value, when assuming $\tilde{c} \approx .7$, and also setting $B \approx B_C \propto 10^{+33}$ so that

$$R \equiv R_0 \cong \left(\frac{\tilde{c} \cdot B^{4/3}}{8 \cdot \pi \cdot \sigma}\right)^{1/3} \tag{12}$$

If we wish to have this of the order of magnitude of a Planck length $l_P$, then the axion domain wall tension must be huge, which is not unexpected. Still though, this pre supposes a minimum value of $B$ which Zhitnitsky[14] set as

$$B_C^{\exp} \sim 10^{20} \tag{13}$$

We need to keep in mind that Zhitnitsky[14] set this parameterization up to account for a dark matter candidate. I am arguing that much of this same concept is useful for setting up an initial condensate of di quark pairs as, separately S-S' in the initial phases of nucleation, with the further assumption that there is an analogy with the so called color super conducting phase (CS) which would permit di quark channels. The problem we are analyzing not only is equivalent to BCS theory electron pairs but can be linked to creating a region of nucleated space in the onset of inflation which has S-S' pairs. The S-S' pairs would have a distance between them proportional to distance mentioned earlier, $R_0$, which would be greater than or equal to the minimum Planck's distance value of $l_P$. The moment one would expect to have deviations from the flat space geometry would closely coincide with Rocky Kolb's model for when degrees of freedom would decrease from over 100 degrees of freedom to roughly ten or less during an abrupt QCD phase transition. The QCD phase transition would be about the time one went from the first to the second potential systems mentioned above.

## V. HOW THIS TIES IN WITH REGARDS TO THE SCHERRER K ESSENCE MODEL RESULTS

We[15] have investigated the role an initial false vacuum procedure with a driven sine Gordon potential plays in the nucleation of a scalar field in inflationary cosmology. Here, we show how that same scalar field blends naturally into the chaotic inflationary cosmology presented by Guth[7] which has its origins in the evolution of nucleation of an electron-positron pair in a de Sitter cosmology. The final results of this model, when $\phi \to \varepsilon^+$, appears congruent with the existence of a region that matches the flat slow roll requirement of $\left|\frac{\partial^2 V}{\partial \phi^2}\right| << H^2$; the negative pressure requirement involving both first and second derivatives of the potential w.r.t. scalar fields divided by the potential itself being very small quantities, where $H$ is the expansion rate that is a requirement of realistic inflation models.[7] This is due to having the potential in question $V \propto \phi^2 \xrightarrow[\phi \to \varepsilon^+]{} V_0 \equiv$ constant for declining scalar values.

We have formed, using Scherrer's argument,[6] a template for evaluating initial conditions to shed light on whether this model universe is radiation-dominated in the beginning or is more in sync with having its dynamics determined by assuming a straight cosmological constant. Our surprising answer is that we do not have conditions for formation of a cosmological constant-dominated era when close to a thin wall approximation of a scalar field of a nucleating universe, but that this is primarily due to an extremely sharp change in slope of the would-be potential field ϕ. The sharpness of this slope, leading to a near delta function behavior for kinematics at the thin wall approximation for the initial conditions of an expanding universe would lead, at a later time, to conditions appropriate for necessary and sufficient cosmological dynamics largely controlled by a cosmological constant when the scalar field itself ceases to be affected by the thin wall approximation but is a general slowly declining slope.

## VI. HOW DARK MATTER TIES IN, USING PURE KINETIC *K* ESSENCE AS DARK MATTER TEMPLATE FOR A NEAR THIN WALL APPROXIMATION OF THE DOMAIN WALL FOR $\phi$

We define k essence as any scalar field with non-canonical kinetic terms. Following Scherrer,[6] we introduce a momentum expression via

$$p = V(\phi) \cdot F(X) \tag{14}$$

where we define the potential in the manner we have stated for our simulation as well as set[13]

$$X = \frac{1}{2} \cdot \nabla_\mu \phi \ \nabla^\mu \phi \tag{15}$$

and use a way to present *F* expanded about its minimum and maximum[6]

$$F = F_0 + F_2 \cdot (X - X_0)^2 \tag{16}$$

where we define $X_0$ via $F_X\big|_{X=X_0} = \frac{dF}{dX}\bigg|_{X=X_0} = 0$, as well as use a density function[6]

$$\rho \equiv V(\phi) \cdot [2 \cdot X \cdot F_X - F] \tag{17}$$

where we find that the potential neatly cancels out of the given equation of state so[6]

$$w \equiv \frac{p}{\rho} \equiv \frac{F}{2 \cdot X \cdot F_X - F} \tag{18}$$

as well as a growth of density perturbations terms factor Garriga and Mukhanov[16] wrote as

$$C_x^2 = \frac{(\partial p / \partial X)}{(\partial \rho / \partial X)} \equiv \frac{F_X}{F_X + 2 \cdot X \cdot F_{XX}} \tag{19}$$

where $F_{XX} \equiv d^2 F / dX^2$, and since we are fairly close to an equilibrium value, we pick a value of X close to an extremal value of $X_0$.[6]

$$X = X_0 + \tilde{\varepsilon}_0 \tag{20}$$

where, when we make an averaging approximation of the value of the potential as very approximately a constant, we may write the equation for the k essence field as taking the form (where we assume $V_\phi \equiv dV(\phi)/d\phi$)

$$(F_X + 2 \cdot X \cdot F_{XX}) \cdot \ddot{\phi} + 3 \cdot H \cdot F_X \cdot \dot{\phi} + (2 \cdot X \cdot F_X - F) \cdot \frac{V_\phi}{V} \equiv 0 \tag{21}$$

as approximately

$$(F_X + 2 \cdot X \cdot F_{XX}) \cdot \ddot{\phi} + 3 \cdot H \cdot F_X \cdot \dot{\phi} \cong 0 \tag{22}$$

which may be re written as[6]

$$(F_X + 2 \cdot X \cdot F_{XX}) \cdot \ddot{X} + 3 \cdot H \cdot F_X \cdot \dot{X} \cong 0 \tag{23}$$

In this situation, this means that we have a very small value for the growth of density pertubations[6]

$$C_S^2 \cong \frac{1}{1+2\cdot(X_0+\tilde{\varepsilon}_0)\cdot(1/\tilde{\varepsilon}_0)} \equiv \frac{1}{1+2\cdot\left(1+\dfrac{X_0}{\cdot\tilde{\varepsilon}_0}\right)} \tag{24}$$

when we can approximate the *kinetic energy* from

$$(\partial_\mu\phi)\cdot(\partial^\mu\phi) \equiv \left(\frac{1}{c}\cdot\frac{\partial\phi}{\partial\cdot t}\right)^2 - (\nabla\phi)^2 \cong -(\nabla\phi)^2 \rightarrow -\left(\frac{d}{dx}\phi\right)^2 \tag{24a}$$

and, if we assume that we are working with a comparatively small contribution w.r.t. time variation but a very large, in many cases, contribution w.r.t. spatial variation of phase

$$|X_0| \approx \frac{1}{2}\cdot\left(\frac{\partial\phi}{\partial x}\right)^2 \gg \tilde{\varepsilon}_0 \tag{24b}$$

$$0 \leq C_S^2 \approx \varepsilon^+ \ll 1 \tag{25}$$

And[6,17]

$$w \equiv \frac{p}{\rho} \cong \frac{-1}{1-4\cdot(X_0+\tilde{\varepsilon}_0)\cdot\left(\dfrac{F_2}{F_0+F_2\cdot(\tilde{\varepsilon}_0)^2}\cdot\tilde{\varepsilon}_0\right)} \approx 0 \tag{26}$$

We get these values for the phase $\phi$ being nearly a box, i.e. the thin wall approximation for *b* being very large in Eq. (4); this is consistent with respect to Eq. (26) main result, with $w \equiv \dfrac{p}{\rho} \cong 0 \Rightarrow$ treating the potential system given by the first potential (modified sine Gordon with small quantum mechanical driving term added) as a semi classical system obeying Eq. (6). This also applies to the formation of S-S' pair formation due to the di quarks as alluded to in Zhitinisky's[14] formulation of QCD balls with an axion wall squeezer having a 'thin wall' character.

When we observed

$$|X_0| \approx \frac{1}{2} \cdot \left(\frac{\partial \phi}{\partial x}\right)^2 \cong \frac{1}{2}\left[\delta_n^2(x+L/2) + \delta_n^2(x-L/2)\right] \qquad (27)$$

with

$$\delta_n(x \pm L/2) \xrightarrow[n \to \infty]{} \delta(x \pm L/2) \qquad (28)$$

as the slope of the S-S' pair approaches a box wall approximation in line with thin wall nucleation of S-S' pairs being in tandem with $b \to$ *larger*. Specifically, in our simulation, we had $b \to 10$ above, rather than go to a pure box style representation of S-S' pairs; this could lead to an unphysical situation with respect to delta functions giving infinite values of infinity, which would force both $C_s^2$ and $w \equiv \frac{p}{\rho}$ to be zero for

$$|X \approx X_0| \cong \frac{1}{2} \cdot \left(\frac{\partial \phi}{\partial x}\right)^2 \to \infty$$ if the ensemble of **S-S'** pairs were represented by a pure thin

wall approximation,[1] i.e., a box. If we adhere to a finite but steep slope convention to modeling both $C_s^2$ and $w \equiv \frac{p}{\rho}$, we get the following: When $b \geq 10$ we obtain the conventional results of

$$w \cong \frac{-1}{1-4\cdot\frac{X_0\cdot\tilde{\varepsilon}_0}{F_2}} \to -1 \tag{29}$$

and recover Scherrer's solution for the speed of sound[6]

$$C_S^2 \approx \frac{1}{1+4\cdot X_0\left(1+\frac{X_0}{2\cdot\tilde{\varepsilon}_0}\right)} \to 0 \tag{30}$$

(If an example $F_2 \to 10^3, \tilde{\varepsilon}_0 \to 10^{-2}, X_0 \to 10^3$). Similarly, we would have if $b \to 3$ in Eq. (5)

$$w \cong \frac{-1}{1-4\cdot\frac{X_0\cdot\tilde{\varepsilon}_0}{F_2}} \to -1 \tag{31}$$

and

$$C_S^2 \approx \frac{1}{1+4\cdot X_0\left(1+\frac{X_0}{2\cdot\tilde{\varepsilon}_0}\right)} \to 1 \tag{32}$$

if $F_2 \to 10^3, \tilde{\varepsilon}_0 \to 10^{-2}$. Furthermore $|X_0| \to$ *a small value*, which for $b \to 3$ in Eq. (5) would lead to $C_S^2 \approx 1$, i.e., when the wall boundary of a S-S' pair is no longer approximated by the thin wall approximation. This eliminates having to represent the initial state as behaving like pure radiation state (as Cardone[17] postulated), i.e., we then recover the cosmological constant. When $|X_0| \approx \frac{1}{2}\cdot\left(\frac{\partial\phi}{\partial x}\right)^2 \gg \tilde{\varepsilon}_0$ no longer holds, we can have a hierarchy of evolution of the universe as being first radiation dominated, then dark matter, and finally dark energy.

$$\text{If } |X \approx X_0| \cong \frac{1}{2} \cdot \left(\frac{\partial \phi}{\partial x}\right)^2 \to \infty,$$ neither limit leads to a physical simulation that makes sense; so, in this problem, we then refer to the contributing slope as always being large but not infinite. We furthermore have, even with $w = -1$

$$C_s^2 \equiv 1 \xrightarrow[b1 \to 3]{} 1 \quad ^3 \tag{33}$$

indicating that the evolution of the magnitude of the phase $\phi \to \varepsilon^+$ corresponds with a reduction of our cosmology from a dark energy dark matter mix to the more standard cosmological constant models used in astrophysics. This coincidently is when the semi classical evaluation involving S-S' di quark pairs breaks down, as given by Eq. (7) and corresponds to the b of Eq. (5) for $\phi \to \varepsilon^+$ being quite small. It also denotes a region where there is a dramatic reduction of the degrees of freedom of the FRW space time metric, as Kolb postulated so that we can then visualize cosmological dynamics being governed by the Einstein constant at the conclusion of the cosmological inflationary period.

## VII. CONCLUSION

Veneziano model[4] gives us a neat prescription of the existence of a Planck's length dimensionality for the initial starting point for the universe via:

$$l_P^2 / \lambda_S^2 \approx \alpha_{GAUGE} \approx e^\phi \tag{34}$$

where the weak coupling region would correspond to where $\phi \ll -1$ and $\lambda_s$ is a so called quanta of length, and $l_P \equiv c \cdot t_P \sim 10^{-33} cm$. As Veneziano implies by his 2nd figure [4], a so called scalar dilaton field with these constraints would have behavior seen by the right hand side of his figure one, with the $V(\phi) \to \varepsilon^+$ but would have no guaranteed false minimum $\phi \to \phi_F < \phi_T$ and no $V(\phi_T) < V(\phi_F)$. The typical string models assume that we have a present equilibrium position in line with strong coupling corresponding to $V(\phi) \to V(\phi_T) \approx \varepsilon^+$ but no model corresponding to potential barrier penetration from a false vacuum state to a true vacuum in line with Coleman's presentation.[3] However, FRW cosmology[18] will in the end imply

$$t_P \sim 10^{-42} \sec onds \Rightarrow size \quad of \quad universe \approx 10^{-2} cm \tag{35}$$

which is still huge for an initial starting point, whereas we manage to in our S-S' 'distance model' to imply a far smaller but still non zero radii for the initial 'universe' in our model.

We find that the above formulation in Eq. (34) is most easily accompanied by the given S-S' di quark pair basis for the scalar field used in this paper, and that it also is consistent with the initial scalar cosmological state evolving toward the dynamics of the cosmological constant via the *k* essence argument built up near the end of this document. Furthermore, we also argue that the semi classical analysis of the initial potential system as given by Eq. (7) and its subsequent collapse is de facto evidence for a phase transition to conditions allowing for CMB to be created at the beginning of inflationary cosmology.

We are fortunate as shown in **Appendix I** that for determining the relative good fit of Eq. (7) that the relative domain walls slope of the initial phase given by Eq. (5) was not terribly significant, for the first potential system, which dove tails with Eq. (1) merely pushing out the domain walls, as a primary effect, for a driven sine Gordon type modeling of false vacuum nucleation. As ,mentioned earlier, this was actually heightened by the extra dimensionality as alluded to by the power law relationship in Eq. (1) making an almost perfect equality between the left and right hand sides of Eq. (7). That the different sides of Eq. (7) in **Appendix I** had varying values, showing different degrees of break down of this relationship for the $2^{nd}$ transitional potential, due to differences in dimensionality and slope of the scalar field as given by Eq. (5) is probably due to this representing the abrupt loss of numbers of degrees of freedom Rocky Kolb has mentioned as part of a phase transition. Needless to say though, as we evolve toward the Einstein cosmological constant era and chaotic inflation, as given by the $3^{rd}$ potential, we should keep in mind very real limits as to the comparative sharpness of the slope of the scalar field as given by Eq. (5)

K essence analysis argues against making *b* in Eq. (5) too large, i.e., if we have a 'perfect' thin wall approximation to our S-S' di quark pairs, we will have the unphysical speed of sound results plus other consequences detailed in the k essence section of the document which we do not want. On the other hand, the semi classical analysis brought up in the section starting with Eq. (7) shows us that a close to the thin wall approximation for S-S' di quark pairs gives an optimal fit for consistency in the potential with the wave functions exhibiting a thin wall approximation 'character'. It is useful to note that our kinetic model can be compared with the very interesting Chimentos[19] purely kinetic k – essence model, with density fluctuation behavior at the initial start of a nucleation process. The model indicate our density function reach $\rho =$ constant after passing through the tunneling barrier as mentioned in our nucleation of a S-S' pair ensemble. This is when the Einstein constant becomes dominant and that the semi classical approximation in Eq. (7) for a domain wall at the time the comparative thin wall approximation S-S' pair ceases to be relevant.

Further developments of this idea would entail more concreted modeling of initial wave functional states than the admittedly very crude start given by Eq. (6). I also am convinced that the quantum fluctuation idea, as referenced by that $\phi^*$ term in Eq. (4a) is a trigger for entropy growth which would be a convenient start for the break down of the scalar field thin wall approximation I used in the initial phases of this document. This would not be materially different from utilizing

$$\delta \cdot E = T \cdot dS \tag{36}$$

as a pre cursor to an energy fluctuation at the beginning of the nucleation process being in tandem with changes in entropy. Note that both Eq. (6) and Eq. (7) were done in the simplest manner possible. This very likely should be re visited, especially if the sort of brane world objects referred to by Trodden et al[9,10] are used in a future calculation for initial nucleation states.

**[Insert figures 1a, 1b, and then figures 2a, 2b *with captions* here]**

# APPENDIX I:

# HOW TO WORK WITH EQUATION 7 FOR MODELING THE EXISTENCE OF SEMI CLASSICAL BEHAVIOR IN AN EARLY UNIVERSE MODEL

For the first potential system, if we set xb=1, xa= - 1, and b = 10. (a sharp slope) for the scalar field boundary we have.

$$\alpha := \frac{.373}{1} \tag{1}$$

This assumes a Gaussian wave functional of

$$\psi(x) := \exp(-\alpha \cdot \phi(x)) \tag{2}$$

As well as a power parameter of

$$\nu := 9 \tag{3}$$

Also, we are using, initially, a phase evolution parameter of

$$\phi(x) := \pi \cdot [\tanh[b \cdot (x - xa)] - \tanh[b \cdot (xb - x)]] \tag{4}$$

The first potential system is re scaled as

$$V1(x) := \frac{1}{2} \cdot (1 - \cos(\phi(x))) - \frac{1}{200} \cdot (\phi(x) - \pi)^2 \tag{5}$$

In addition, the following is used as a rescaling of the inner product

$$c1 := \frac{1}{\displaystyle\int_{-30}^{30} (\exp(-\alpha \cdot \phi(x)))^2 \cdot \frac{\pi^3}{3} \cdot x^5 \, dx} \tag{6}$$

$$c2 := \int_{-30}^{30} (\exp(-\alpha \cdot \phi(x)))^2 \cdot \frac{\pi^3}{3} \cdot x^5 \cdot (V1(x))^\nu \cdot |c1| \, dx \qquad (7)$$

$$c3 := \left[ \int_{-30}^{30} (\exp(-\alpha \cdot \phi(x)))^2 \cdot \frac{\pi^3}{3} \cdot x^5 \cdot V1(x) \cdot |c1| \, dx \right]^\nu \qquad (8)$$

$$c3b := \frac{c2}{c3} \qquad (9)$$

Here,

$$C3b = .999 \qquad (9a)$$

For the 2$^{nd}$ potential system, if we assume a sharp slope, i.e. b1 = b = 10, and

$$V2(x) := \frac{1}{2} \cdot \frac{(\phi a(x))^2}{1 + .000001 \cdot (\phi a(x))^3} \qquad (10)$$

If

$$\phi a(x) := \pi \cdot [\tanh[b1 \cdot (x - xa)] - \tanh[b1 \cdot (xb - x)]] \qquad (11)$$

and a modification of the 'Gaussian width' to be

$$\alpha 1 := \frac{.373}{30} \qquad (12)$$

We do specify a denominator, due to a normalization contribution we write as

$$c1a := \frac{1}{\int_{-30}^{30} (\exp(-\alpha 1 \cdot \phi a(x)))^2 \cdot \frac{\pi^3}{3} \cdot x^5 \, dx} \qquad (13)$$

$$c4 := \int_{-30}^{30} (\exp(-\alpha 1 \cdot \phi a(x)))^2 \cdot \frac{\pi^3}{3} \cdot x^5 \cdot (V2(x))^\nu \cdot |c1a| \, dx$$

(14)

In addition:

$$c5 := \left[ \int_{-30}^{30} (\exp(-\alpha \cdot \phi a(x)))^2 \cdot \frac{\pi^3}{3} \cdot x^5 \cdot V2(x) \cdot |c1a| \, dx \right]^\nu$$

(15)

We then use a ratio of

$$c5b := \frac{c4}{c5}$$

(16)

Here, when one has the six dimensions, plus the thin wall approximation:

$$C5b = 2.926E\text{-}3$$

(17)

When one has three dimensions, plus the thin wall approximation

$$c6 := \int_{-30}^{30} (\exp(-\alpha 1 \cdot \phi a(x)))^2 \cdot \frac{\pi^1}{.25} \cdot x^2 \cdot (V2(x))^\nu \cdot |c1b| \, dx$$

(18)

$$c7 := \left[ \int_{-30}^{30} (\exp(-\alpha \cdot \phi(x)))^2 \cdot \frac{\pi^1}{.25} \cdot x^2 \cdot V2(x) \cdot |c1b| \, dx \right]^\nu$$

(19)

$$c7b := \frac{c6}{c7}$$

(20)

This leads to

$$c7b = .019 \tag{21}$$

When one has the thin wall approximation removed, via b1 = 1.5, one does not see a difference in the ratios obtained.

For the 3$^{rd}$ potential system, which is intermediate between the 1$^{st}$ and 2$^{nd}$ potentials if the b1 = b = 10 value is used, one obtains for when we have six dimensions

$$\alpha 1 := \frac{.373}{6} \tag{22}$$

As well as

$$V2(x) := \frac{1}{2} \cdot \frac{(\phi a(x))^2}{1 + .5 \cdot (\phi a(x))^3} \tag{23}$$

(When we have six dimensions)

$$C5b = 0.024 \tag{24}$$

(When we have three dimensions)

$$C7b = .016 \tag{25}$$

So, then one has C5b = .024, and C7b = .016 in the thin wall approximation

When b1 = 3 (non thin wall approximation)

$$C5b = .027 \tag{26}$$

(**Six dimensions**)

$$C7b = .02 \tag{27}$$

(**three dimensions**)

Summarizing, if

$$V1(x) := \frac{1}{2} \cdot (1 - \cos(\phi(x))) - \frac{1}{200} \cdot (\phi(x) - \pi)^2 \qquad = V1 \qquad (28)$$

$$V2(x) := \frac{1}{2} \cdot \frac{(\phi a(x))^2}{1 + .000001 \cdot (\phi a(x))^3} \qquad = V3 \qquad (29)$$

$$V2(x) := \frac{1}{2} \cdot \frac{(\phi a(x))^2}{1 + .5 \cdot (\phi a(x))^3} \qquad = V2 \qquad (30)$$

One finally obtains the following results, as summarized below

|           | b=b1 = 10      | b1 = 3       | b1 = 1            |
|-----------|----------------|--------------|-------------------|
| V1 ( 6 dim) | C3b = .999     | No data      | No data           |
| V3 ( 6 dim) | C5b = 2.926E-3 | No data      | C5b =  same value |
| V3 ( 3 dim) | C7b = .019     | No data      | C7b =  same value |
| V2( 6 dim)  | C5b =  .027    | C5b =  .024  | No data           |
| V2 ( 3 dim) | C7b =  .02     | C7b =  .016  | No data           |

# Figure captions

**Fig 1a,b:** Evolution of the phase from a thin wall approximation to a more nuanced thicker wall approximation with increasing L between S-S' instanton components. The 'height' drops and the 'width' L increases correspond to a de evolution of the thin wall approximation. This is in tandem with a collapse of an initial nucleating 'potential' system to the standard chaotic scalar $\phi^2$ potential system of Guth. As the 'hill' flattens, and the thin wall approximation dissipates, the physical system approaches standard cosmological constant behavior.

**Fig 2a,b**: As the walls of the S-S' pair approach the thin wall approximation, a normalized distance, $L = 9 \to \mathbf{L = 6} \to L = 3$, approaches delta function behavior at the boundaries of the new nucleating phase. As $L$ increases, the delta function behavior subsides dramatically. Here, the $L = 9 \Leftrightarrow$ conditions approaching a cosmological constant. $\mathbf{L} = 6 \Leftrightarrow$ conditions reflecting Scherrer's dark energy-dark matter mix. $L = 3 \Leftrightarrow$ approaching unphysical delta function contributions due to a pure thin wall model.

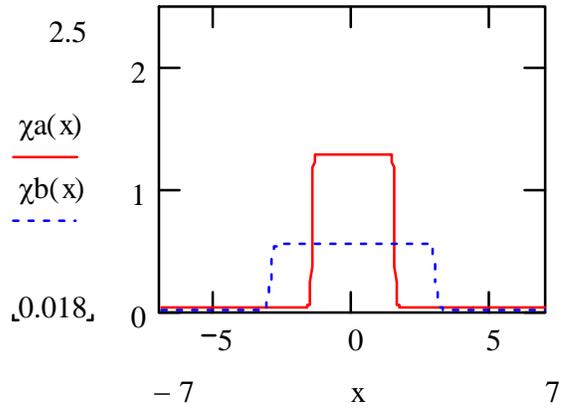

**Figure 1a**

**Beckwith**

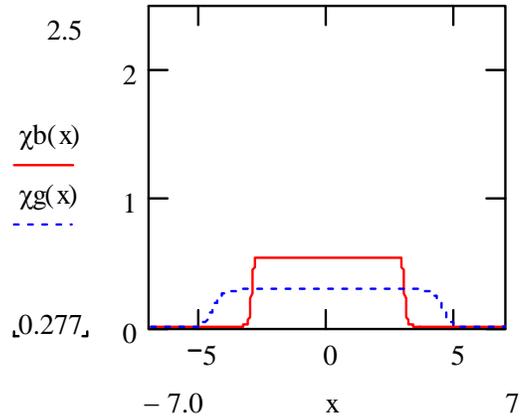

**Figure 1b**

**Beckwith**